\documentstyle[aaspptwo]{article}
\tighten
\slugcomment{submitted to ApJ {\em Letters\/}}

\newcommand{\beq}{\begin{equation}}
\newcommand{\eeq}{\end{equation}}
\newcommand{\beqa}{\begin{eqnarray}}
\newcommand{\eeqa}{\end{eqnarray}}

\newcommand{\msun}{\hbox {$M_{\odot}$ }}

\newcommand{\dv}{$\Delta {\rm V} ({\rm TO} - {\rm HB})$ }

\newcommand{\dvtwo}{\Delta {\rm V}}

\newcommand{\ea}{{\it et al. }}

\newcommand{\feh}{\hbox{$ [{\rm Fe}/{\rm H}]$ } }
\newcommand{\kmsmpc}{\hbox{$ {\rm km}\, {\rm s}^{-1}\, {\rm Mpc}^{-1}$ }}

\begin{document}

\title{Absolute Ages of Globular Clusters and the Age of the Universe}

\author{Brian Chaboyer}

\affil{Canadian Institute for Theoretical Astrophysics, \\
60 St. George Street,  Toronto, Ontario, Canada  M5S 1A7\\
E-Mail: chaboyer@cita.utoronto.ca}

\begin{abstract}
The main sequence turnoff luminosity is the best stellar `clock' which
can be used to determine the absolute ages of globular clusters.  This
is due to the fact that it is generally assumed that the luminosity
and lifetimes of main sequence globular cluster stars are independent
of the properties of stellar convection and atmospheres, two areas of
stellar evolution which are poorly understood.  Several possible
sources of error in this stellar clock are discussed, and isochrones
are constructed using a variety of different physical assumptions.
The mean age of the oldest globular clusters are determined from these
isochrones and it is found that the uncertainties in the input physics
can lead to changes in the derived age of $\pm 15\%$.  Surprisingly
the largest source of error is the mixing length theory of convection.
It is well known that uncertainties in the distance scale and chemical
composition of globular cluster stars lead to changes of order $\sim
22\%$ in the determination of absolute ages.  Combining the various
sources of error, the absolute age of the oldest globular clusters are
found to lie in the range 11 --- 21 Gyr.  This is meant to be a total
theoretical range.  For the standard inflationary model ($\Omega = 1,
 ~ \Lambda = 0$), a minimum age of the universe of 11 Gyr requires $H_o
\la 60~\kmsmpc$.
\end{abstract}

\keywords{early universe -- distance scale -- globular clusters:
general -- stars: interiors -- stars: evolution}
\vspace*{20pt}
\begin{center}
{\large submitted to  {\it Astrophysical Journal Letters\/}}
\end{center}
\vspace*{-18cm}
\hspace*{14cm}
CITA--94--52

\section{Introduction}
Due to their low metallicity and nearly spherical distribution about
the galactic centre, it is evident that the Galactic globular clusters
were among the first objects to form in our Galaxy.  As such, an
accurate determination of the age of the Galactic globular clusters
provides a reasonable estimate for the age of the universe.  This fact
has long been recognized, and determination of globular cluster ages
has a rich history (see Demarque, Deliyannis \& Sarajedini 1991 for a
brief summary).  One of the great uncertainties in stellar models is
the treatment of convection.  Due to this uncertainty, stellar models
are not reliable in regions where convection is important.  This
implies the outer layers of the model for the low mass main sequence
stars which make up globular clusters. The cores of these stars (where
the nuclear energy generation occurs) are not convective, and so the
modeled stellar lifetimes and luminosities are assumed to be
unaffected by convection.  For this reason, the luminosity of the main
sequence turnoff (MSTO) is the best stellar clock with which to
determine globular cluster ages (e.g.~Sandage 1970).  Globular
clusters are typically found to have ages of $\sim 15$ Gyr
(Demarque \ea 1991; Chaboyer,
Sarajedini \& Demarque 1992; Bergbusch \& VandenBerg 1992; Salaris,
Chieffi \& Straniero 1993).  This has important consequences for
cosmology.  For example if $\Omega = 1, ~ \Lambda = 0$, a minimum
age of the universe of 15 Gyr requires $H_o \la 44~\kmsmpc$, a value
below recent determinations (Freedman \ea 1994; Pierce \ea 1994;
Riess, Press, \& Kirshner 1994; Tammann \& Sandage 1994).  However,
before making any judgments regarding the validity of a particular
cosmological model based on determinations of $H_o$ and globular
cluster ages, one must first understand the associated errors.

Numerous authors have pointed out that the inferred ages of globular
clusters determined by the MSTO are quite sensitive to uncertainties
in the distance to globular clusters, and the chemical composition of
globular cluster stars (Demarque 1980; Rood 1990; VandenBerg 1990;
Renzini 1991).  However, there has not been a detailed study of errors
arising from uncertainties in the physics which are assumed in the
stellar models.  This {\it Letter\/} will examine the possible sources
of error in stellar models, and determine how these errors affect the
absolute age estimates for globular clusters.

\section{Isochrone Construction and Age Determination}
For each set of assumed input physics, a series of stellar models with
masses ranging from $M=0.5~\msun$ to $1.0~\msun$ (in $0.05\,\msun$
increments) were evolved from the zero-age main sequence to the giant
branch using the Yale stellar evolution code (Guenther \ea 1992).  In
order to span the observed range in globular clusters metallicities,
the models were evolved with $Z = 6\times 10^{-5}$, $2\times 10^{-4}$,
$6\times 10^{-4}$, $2\times 10^{-3}$, $4\times 10^{-3}$, and $7\times
10^{-3}$.  The following physics was assumed in our standard models:
pp reaction rates as tabulated by Bahcall \& Pinsonneault (1992); CNO
reaction rates as tabulated by Bahcall (1989); high temperature
opacities from Iglesias \& Rogers (1991); low temperature opacities
($T < 10^4$K) from Kurucz (1991); surface boundary conditions are
determined using a grey atmosphere; for temperatures above $10^6$ K, a
relativistic degenerate, fully ionized equation of state is used;
below $10^6$ K, the single ionization of $^1$H , the first ionization
of the metals and both ionizations of $^4$He are taken into account
via the Saha equation.  The standard models use a solar calibrated
mixing length ($\alpha = 1.7$), $^4$He diffusion is ignored and a
$^4$He abundance of $Y=0.23$.  This is the same set of physics adopted
as in the new Yale isochrones (Chaboyer \ea 1995).

Observations of globular cluster and halo stars indicate that the
$\alpha$-elements are enhanced over their solar value. This fact was
taken into account by modifying the relationship between \feh and $Z$
(Chaboyer \ea 1992; and Salaris \ea 1993).  For our standard case, the
the chosen $Z$ values correspond to $\feh = -2.8; ~ -2.3; ~-1.8;
 ~-1.3$ all with $[\alpha/{\rm Fe}] = +0.40$; $\feh = -0.9$,
$[\alpha/{\rm Fe}] = +0.30$; and $\feh = -0.6$, $[\alpha/{\rm Fe}] =
+0.25$.  These $[\alpha/{\rm Fe}]$ values have been chosen to be in
agreement with the observations (e.g.~Lambert 1989; Brown \&
Wallerstein 1992; Tomkin \ea 1992; King 1993).

Isochrones were constructed by interpolating among the stellar
evolutionary tracks to a given age.  For each set of input tracks,
isochrones were constructed with ages from 9 -- 22 Gyr, in 1 Gyr
increments.  The colour transformation of Green, Demarque \& King
(1987) was used to convert from theoretical luminosities and colours
to observed magnitudes and colours.  The effect of changing various of
the above assumption was examined by evolving sets of stellar models
using different assumptions for the input physics.

In order to compare the isochrones to the observations, one must
determine the distance to the globular clusters.  This can be done
using main sequence fitting, or using the observed absolute magnitude
of the RR Lyr stars.  Main sequence fitting requires accurate
parallaxes of a number of metal-poor stars for an accurate distance
determination.  Unfortunately, such as sample does not exist at
present, and so the absolute magnitude of the RR Lyr stars was adopted
as our distance modulus.  This allows one to determine the ages using
the difference in magnitude between the main sequence turn-off, and
the horizontal branch (in the the RR Lyr instability strip).  This age
determination technique is commonly referred to as \dv ~and has the
advantage of being independent of cluster reddening.  Unfortunately,
RR Lyr stars have convective cores, and so the theoretical calibration
of their absolute magnitude is subject to large uncertainties due to
the treatment of convection. However, there are several independent
observational methods which can be used to determine the absolute
magnitude of the RR Lyr stars: (1) Baade-Wesslink and infrared flux
methods (Carney, Storm \& Jones 1992; Skillen \ea 1993) ; (2)
statistical parallax observations (Layden \ea 1995); (3) the
Oosterhoff period-shift effect (Sandage 1993); and (4) determining the
apparent magnitude of RR Lyrs in the LMC and using the known distance
of the LMC to determine the absolute magnitudes (Walker 1992).  These
methods have found that the absolute magnitude of the RR Lyr stars is
given by:
\beq
M_v({\rm RR Lyr}) = \alpha \, \feh + \beta.
\label{mvrr}
\eeq
Determinations of the slope, $\alpha$ vary from $0.15$ to $0.30$.
Although the slope with metallicity is important in determining the
relative ages of the globular clusters, it is not important for
deriving the mean absolute age of a number of globular clusters.  In
this study, $\alpha = 0.22$ was chosen.  The absolute age depends
sensitively upon the zero-point, $\beta$.  Galactic determinations of
this zero-point agree to within $\sim 0.15$ mag.  However, the LMC RR
Lyr calibration is $0.25$ mag brighter (Walker 1992).  Ages will be
derived using both the Walker (1992) and Layden \ea (1995) Galactic
zero-point which differ by 0.25 mag, and so represent the maximum
uncertainty in determining the distance.

The MSTO magnitudes are combined with equation (\ref{mvrr}) to
generate a grid of age (in Gyr, $t_9$), \dv ~and \feh which  was
modeled using an equation of the form
\beqa
t_9 & = &a_o + a_1\dvtwo + a_2\dvtwo ^2 + a_3\feh \nonumber \\
     &+ &a_4\feh ^2 + a_5\dvtwo \feh.
\label{age}
\eeqa
The rms residuals of the points about the fit were typically 0.15 Gyr.
Observationally, it is difficult to determine the MSTO luminosity, as
the stars in this region of the colour-magnitude diagram form a nearly
vertical strip.  Typical errors in determining \dv are $\pm 0.14$ mag,
which translates into an error in the derived age ($\pm 15\%$).  To
avoid this difficulty, one may determine the mean age for a large
number of globular clusters.  However, it has become increasingly
clear that some globular clusters are significantly younger than the
mean.  A sample of 24 globular clusters which have nearly the same age
and are unequivocoly old has been selected on the basis of their HB
morphology (Zinn 1993) and/or the $\Delta (B - V)$ precision age
ranking technique (Sarajedini \& Demarque 1990; VandenBerg, Bolte \&
Stetson 1990; Buonanno \ea 1994).  The observed \dv ~are taken from
the compilation of Chaboyer \ea (1992). The cluster metallicities are
taken from Zinn \& West (1984).  Equation (\ref{age}) was used to
determine the ages of of these clusters with the error in the age
being derived from the observed errors in \dv and \feh.  The mean age
was then formed as a weighted sum of the 24 ages.  The error in the
mean age was found to be $\pm 3\%$ ($1\,\sigma)$.

\section{Results}
The main results of the {\it Letter\/} are summarized in Table
\ref{tab1} where the mean age of the oldest globular clusters is
tabulated under a variety of assumptions.  Each case will be discussed
in turn.
\begin{planotable}{clcrcr}
\tablewidth{38pc}
\tablecaption{TABLE 1}
\tablecaption{Mean Age of the Oldest Globular Clusters\tablenotemark{a} }
\tablehead{
& & \colhead{Layden\tablenotemark{b}}
& & \colhead{Walker\tablenotemark{c}}\nl
\colhead{Case}&
\colhead{Description}&
\colhead{Age (Gyr)}&
\colhead{Change}&
\colhead{Age (Gyr)}&
\colhead{Change}
}
\startdata
A & {\bf Standard}\tablenotemark{d}
& $\bf 18.2\pm0.6$ & \multicolumn{1}{c}{\bf ---}
& $\bf 14.2\pm 0.5$ & \multicolumn{1}{c}{\bf ---}\nl
\tableline
\multicolumn{6}{c}{\bf Changed Physics}\nl
\tableline
B1 & Very Low Mixing length ($\alpha = 1.0$) & 16.7 &$ -8.2 \%$& 13.4
&$ -5.6 \%$\nl
B2 &  Low Mixing length ($\alpha = 1.5$) & 17.9 &$ -1.6 \%$& 14.0
&$ -1.4 \%$\nl
B3 & High Mixing length ($\alpha = 2.0$) &18.7 &$ +2.7 \%$& 14.4
& $  +1.4 \%$\nl
B4 & Very High Mixing length ($\alpha = 3.0$) & 20.1 &$ +10.4 \%$&15.4
& $  +8.5 \%$\nl
C & Kurucz color transformation & 17.2 &$ -5.5 \%$& 13.4& $ -5.6\%$\nl
D & Kurucz model atmospheres & 18.1 &$ -0.5 \%$& 14.1& $ -0.7 \%$\nl
E & Cox \& Stewart low temperature opacities & 18.4 &$  +1.1 \%$
& 14.2& $  0.0 \%$\nl
F & $0.30~H_p$ adiabatic overshoot at base of scz & 18.2 &$ 0.0 \%$
& 14.1& $ -0.7 \%$\nl
G & LAOL high temperature opacities & 18.2 &$ 0.0 \%$& 14.3& $0.7\%$\nl
H1 &$3\,\sigma$ decrease in nuclear reaction rates  &19.1&$  +4.9
\%$&14.8 & $  +4.2 \%$\nl
H2 &$2\,\sigma$ decrease in nuclear reaction rates&18.6&$  +2.2
\%$&14.5& $  +2.1 \%$\nl
H3 &$2\,\sigma$ increase in nuclear reaction rates &17.9&$ -1.6
\%$&13.9& $ -2.1 \%$\nl
H4 &$3\,\sigma$ increase in nuclear reaction rates &17.8&$ -2.2
\%$&13.8& $ -2.8 \%$\nl
J & Debyre-H\"{u}ckel EOS & 17.0 &$ -6.6 \%$& 13.3& $ -6.3 \%$\nl
K & $^4$He Diffusion & 16.9 &$ -7.1 \%$& 13.1& $ -7.7 \%$\nl
\tableline
\multicolumn{6}{c}{\bf Changed Abundances}\nl
\tableline
L1 & low $^4$He abundance ($Y=0.20$) & 19.1 &$  +4.9 \%$& 14.9& $+4.9 \%$\nl
L2 & high $^4$He abundance ($Y=0.26$) & 17.5 &$ -3.8 \%$& 13.5& $-4.9 \%$\nl
M1 & GC $[\rm Fe/H]$ decreased by $0.10$ dex & 18.4 &$  +1.1 \%$& 14.3
& $  +0.7 \%$\nl
M2 & GC $[\rm Fe/H]$ increased by $0.10$ dex & 18.1 &$ -0.5 \%$& 14.0
& $ -1.4 \%$\nl
N1 & $[\alpha/{\rm Fe}] = +0.6$  & 17.1 &$ -6.0 \%$& 13.2& $-7.0\%$\nl
N2 & $[\alpha/{\rm Fe}] = +0.2$  & 19.4 &$  +6.6 \%$& 15.1&$+6.3 \%$\nl
\tablenotetext{a}{Based on their HB morphology (Zinn 1993) and/or
the $\Delta (B - V)$ age precision age ranking technique.  There are
24 clusters in this group with measured \dv values.  The clusters
span a range in metallicity of $-0.89 \le [{\rm Fe/H}] \le -2.41$.}
\tablenotetext{b}{Ages derived using the Layden \ea (1995) RR Lyr
distance scale}
\tablenotetext{c}{Ages derived using the Walker (1992) RR Lyr distance scale}
\tablenotetext{d}{The input physics and composition for this model is
discussed in \S 2.}
\label{tab1}
\end{planotable}

{\it Case A\/} is the standard one, using the assumptions described in
the previous section.  Note that the uncertainty in the distance
modulus of 0.25 mag (Layden vs. Walker age) gives rise to a 4 Gyr
change in the derived age.  This fact has been noted by many authors
(eg. Renzini 1991).  It is clear that refining the distance estimate
to Galactic globular clusters will play a key role in reducing the
uncertainty in the derived ages.

{\it Case B\/} examines the effect of changes in the mixing length,
which is used to characterize the transport of energy by convection in
the outer layers of the star.  Small departures from the solar
calibrated mixing length ($\alpha = 1.7$) do not alter the derived age
(cases B2 and B3 with $\alpha = 1.5$ and $2.0$ change the age by less
than 3\%).  This is to be expected, as it is commonly assumed that the
MSTO luminosity is independent of the mixing length.  However, if
large or small values of the mixing length are considered (cases B1
and B4 with $\alpha = 1.0$ and $3.0$), then the derived age of the
globular clusters can change by up to 10\%.  This is due to the fact
that changing the mixing length changes the shape of the stellar
evolutionary models, and hence the isochrones.  Thus, the luminosity
of the bluest point on the isochrone can be altered, even though the
age-luminosity relationship is relatively unaffected by changes in the
mixing length.  We conclude that uncertainties in how to treat
convection in stellar models leads to a maximum uncertainty of 10\% in
globular cluster ages derived using the MSTO luminosity.  This is one
of the largest sources of error the age estimates.  The $\alpha = 1.0$
and $\alpha = 3.0$ isochrones do not match observed globular cluster
colour-magnitude diagrams.  Thus, one would be tempted to state that
these isochrones (and the associated large uncertainty in globular
cluster ages) are ruled out by the observations.  However, there are
other uncertainties that can strongly affect the colour of the models
(surface boundary conditions, opacities, colour transformation).  In
addition, there is no compelling reason that the mixing length should be
the same for the Sun and globular cluster stars, or even have the same
value on the main sequence and the giant branch.  For these reasons,
one should not rule out the $\alpha = 1.0$ and $\alpha = 3.0$.

{\it Case C} considers the effect of using the Kurucz (1992)
colour calibration to convert from the modeled luminosities and
temperatures to observed colours and temperatures.  Use of the Kurucz
(1992) colour calibration reduces the derived age by $5\%$.  Determining
colour calibrations is extremely difficult, and comparing colour
calibrations which have been independently derived gives a good
estimate for the error involved in the process.

{\it Case D} uses the Kurucz (1992) model atmospheres for the surface
boundary conditions (the standard case using a grey atmosphere).  The
change is the derived age is very small ($< 0.6\%$), and so
uncertainties in surface boundary conditions used in stellar models
have little effect on our estimate for the absolute age of the
globular clusters.

{\it Case E} uses the Cox \& Stewart (1970) opacities for $T < 10^4$
K as opposed to the Kurucz (1991) opacities used in the standard case.
This change has very little effect on the derived ages, as is to be
expected from the fact that only the surface layers of the star are
affected by changes in the low temperature opacities.

{\it Case F} includes the effect of a rather large (0.30 pressure
scale heights) adiabatic overshoot layer at the base of the surface
convection zone.  This has very little effect on the derived age.
Together, cases D, E, and F demonstrate that the physical conditions
in the core of the star are unaffected by changes in the outer layers
of the star.

{\it Case G} uses the LAOL opacities of Huebner \ea (1977) for the
interior of the model ($T>10^4$K).  The derived age is very similar to
the standard model, which uses the opacities of Iglesias \& Rogers
(1991).  Thus, uncertainties in the calculation of opacities have
little effect on globular cluster ages derived using the MSTO
luminosity.

{\it Case H} investigates the effects of changing the nuclear reaction
rates.  Even changing all of the nuclear reaction rates by $3\,\sigma$
from their tabulate values changes the derived age by less than $5\%$.
Changing the reaction rates by $2\,\sigma$ is perhaps a more
more realistic assessment of the possible error, and indicates that the
possible errors in the nuclear reaction rates have a negligible effect
($\sim 2\%$) on the derived age.  For these low mass, low metallicity
stars, the key nuclear energy generating reaction is the pp chain, and the
small change in the derived age due to changing the nuclear reactions
is a reflection of the fact that the pp cross section is well
understood on theoretical grounds (Bahcall 1989).

{\it Case J} includes the effects of the Debyre-H\"{u}ckel correction
to the equation of state (which takes into account the Coulomb forces,
see Guenther \ea 1992 for a description of its implementation).  It
leads to a decrease in the derived age of $6.5\%$.  However, the
Debyre-H\"{u}ckel correction is not the only non-ideal gas effect
which should be included in a realistic equation of state.  Rogers
(1981; 1986) discusses various corrections to the equation of state in
stellar plasmas, and finds that other effects may be as important (and
have the opposite sign) as the Debyre-H\"{u}ckel.  For this reason, it
is not clear that including the Debyre-H\"{u}ckel correction results
in an equation of state which better reflects the true equation of
state. We conclude that uncertainties in the equation of state can lead to
changes in the derived age of the oldest globular clusters of
$\sim \pm 4\%$.

{\it Case K} includes the effect of the diffusion of $^4$He relative
to $^1$H.  For this study, the diffusion coefficients of Michaud \&
Proffitt (1993) were used and the age of the oldest globular clusters
is reduced by $7\%$. The effects of diffusion on the ages of
globular clusters have been studied by numerous authors
(e.g. Noerdlinger \& Arigo 1980; Proffitt \& VandenBerg 1991;
Chaboyer \ea 1992), and our results are compatible with the later
studies.  However, as pointed out by Chaboyer \& Demarque (1994),
stellar models which only include diffusion are unable to explain the
observed plateau in Li abundances which occurs in halo stars.  For
this reason, it is unclear if diffusion actually occurs in stars,
but it does remain a possible source of error in the stellar models.

{\it Cases L -- N} demonstrate the effects that the uncertainty in the
true chemical composition of globular cluster stars have on the
derived age estimate.  This effect has been studied by others
(e.g. Demarque 1980; VandenBerg 1990; Renzini 1991), and so
is only briefly discussed here.  Changing the $^4$He abundance or an
overall shift in the \feh scale for globular clusters has
little effect on the derived age.  However, the uncertainty in the
true abundance of the $\alpha$-capture elements (and oxygen in particular)
leads to changes in the derived age of the oldest globular cluster of
$\pm 7\%$.  Deriving oxygen abundances from observations is subject to a
number of systematic errors, and a range of $[{\rm O/Fe}] = +0.2$ to
0.6 dex which was used in the calculations is indicative of the
scatter found between different observers (see King 1993).

Although not tabulated in Table 1, a number of cases have been
constructed which combine several of the above effects and it was
found that the percent change in the derived age is well approximated by
adding up the percentage change for the individual cases considered.

\section{Discussion}
The mean age for the oldest globular globular clusters derived using
the MSTO luminosity is tabulated in Table \ref{tab1} for a variety of
different physical assumptions.  Aside from the 22\% change in the
derived age which arises from determining the true distance to the
cluster (the `Layden' and `Walker' ages in Table \ref{tab1}), the
largest single change in the age determination ($\pm 10\%$) is due to
our poor understanding on how to model convection in stellar evolution
models.  This is exemplified by case B which considered different
values of the mixing length.  This is a somewhat surprising result, as
it is commonly assumed that ages derived using the MSTO luminosity are
insensitive to uncertainties in our knowledge of convection.  Other
variables which can lead to large changes in the derived age include
the colour transformation (case C, 5\%), the equation of state (case
J, 7\%) and whether or not to include helium diffusion (case K, 7\%).
This is comparable to the error in the age due to the uncertain
$\alpha$-element composition (case N, $\pm 7\%$).  It is impossible to
quantify a $1\,\sigma$ error in the derived age of the globular
clusters, as many of the sources of error are systematic.  However,
one can conclude that even if the distance and compositions of the
globular clusters were known exactly, the total uncertainty in the
age estimate would be $\pm 15\%$.

The effects of mass loss or rapidly rotating cores has not been
considered in this work.  Shi (1994) found that the age reduction due
to mass loss was constrained by the observations to be less than $\sim
1$ Gyr.  The effect of rapidly rotating cores on stellar models and
isochrones has been studied by Deliyannis, Demarque \& Pinsonneault
(1989) and Chaboyer \& Demarque (1994), who found that rotation has
virtually no effect on the derived ages.  In light of these results,
and of Table \ref{tab1}, one may conclude that the true age of
globular clusters lies in the range 11 -- 21 Gyr.  In the context of
cosmology, it is the lower limit which is of most interest.  To reach
this lower limit a number of non-standard assumptions must be made:
$^4$He diffusion occurs in stars; the Debyre-H\"{u}ckel correction
should be included in the equation of state; mass loss is occurring in
main sequence globular cluster stars and the galactic calibration of
the RR Lyr absolute magnitudes must be in error by 0.25 magnitudes.
Although it is possible that a few of these effects are true, it would
seem unlikely that they all occur.  Based on this, a more realistic
assessment of the globular cluster ages would be 13 -- 17 Gyr.

A lower limit of 11 or 13 Gyr to the age of the universe imposes
strong constraints on the standard inflation model ($\Omega = 1$,
$\Lambda = 0$).  A minimum age of 11 Gyr requires $H_o \la
52~\kmsmpc$, while 13 Gyr requires $H_o \la 60~\kmsmpc$.  Recent
measurements of Cepheids in Virgo by CFHT (Pierce \ea 1994) and HST
(Freedman \ea 1994) find $H_o = 87\pm7$ and $80\pm 17$ respectively.
If the central value remains the same, as the error in the
determination of $H_o$ is reduced, an $\Omega = 1$, $\Lambda = 0$
universe would be ruled out.  A low density $\Omega = 0.1$, $\Lambda
= 0$ universe is compatible with an age of 11 Gyr and $H_o \simeq
80~\kmsmpc$.

\acknowledgments
I would like to thank A.~Layden for providing me with his RR Lyr
calibration and R.~Zinn who shared with me his list of old halo
clusters, both in advance of publication.  I am grateful to
P.~Demarque and R.~Malaney for their comments on an early draft of
this paper.

\appendix

\end{document}